# Contour map of estimation error for Expected Shortfall

**I. Kondor[1,2], F. Caccioli[3], G. Papp[4] and M. Marsili[5]**

[1]Parmenides Foundation, Pullach, Germany
[2]Department of Investment and Corporate Finance, Corvinus University of Budapest, Budapest, Hungary
[3]Department of Computer Science, University College London, London, United Kingdom
[4]Eötvös Loránd University, Institute for Physics, Budapest, Hungary
[5]Abdus Salam International Centre for Theoretical Physics, Trieste, Italy

**Abstract**

The contour map of estimation error of Expected Shortfall (ES) is constructed. It allows one to quantitatively determine the sample size (the length of the time series) required by the optimization under ES of large institutional portfolios for a given size of the portfolio, at a given confidence level and a given estimation error.

ES is on its way to becoming the new regulatory market risk measure [1]. Even though the primary purpose of ES will be to characterize the risk in an institution's existing portfolio, banks will have to optimize their investment and trading activities under the constraints of ES. This is analogous to the classical portfolio selection problem, with ES replacing variance as the cost function.

A critically important feature of ES (shared with all the other downside risk measures, including VaR) is that its *historical estimate* is calculated on the basis of the 1% (or, in the proposed new regulation, the 2.5%) worst outcomes. This means that most of the data in a time series have to be discarded, and only a small fraction is used to carry out the optimization. As a result, the procedure will be fragile, the estimates obtained for the portfolio weights will strongly fluctuate from sample to sample, and the estimation error will be large. This is a well known problem that one may try to resolve by regularization borrowed from high dimensional statistics [2]. We have discussed the instability of ES in a series of papers [3-10], and studied the effects of various regularizers by which one may hope to rein in the fluctuations [7,8,10].

In the above works we mainly discussed the *phase boundary*, the line on the *N/T* vs. α plane along which the (unregularized) estimation error diverges, and beyond which the optimization problem does not have a solution. (*N* is the number of different risky assets in the portfolio, *T* is the length of the time series, and α is the confidence level.) However, the papers just cited contain information not only for the phase boundary, but implicitly also for the complete *countour map* of ES, that is the set of curves along which the estimation error has a given finite value.

The purpose of this note is to display these curves and show by their help that at the high confidence level envisaged by regulation and for an acceptable level of estimation error for the risk of the optimal portfolio, the sample size *T* must exceed the dimension *N* of the portfolio by several factors of ten, a requirement evidently hard if not impossible to satisfy for any realistic values of *N* and *T*.

The contour map of the estimation error for a large portfolio of i.i.d. Gaussian assets is shown in Fig.1. (Correlated and non-Gaussian asset distributions will be briefly discussed later in this note.) We assume that both the number *N* of different assets in the portfolio and the sample size *T* (the length of the available time series for these assets) are large. Then the estimation error will only depend on the ratio *N/T*, and on the confidence level α. Fig.1 displays the set of curves on the *N/T* – α plane, along which the estimation error, measured in terms of the quantity Δ, is constant.

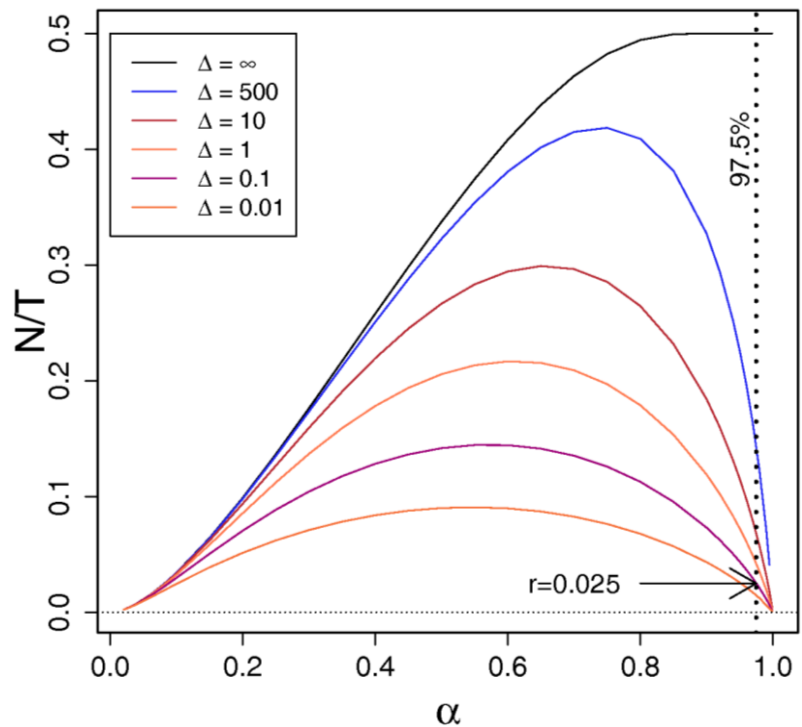

**Fig.1:** Contour map of the estimation error Δ for ES. Δ is constant along these curves. The curves lying above each other correspond to higher and higher values of Δ, up to the uppermost black curve along which the estimation error diverges, and beyond which ES cannot be optimized. At a given confidence level α one can read off from these curves the ratio *N/T* that is required to achieve an acceptable estimation error. For example, at α = 97.5% (the confidence level proposed by the new regulation and shown by the vertical dotted line in the figure) the vertical coordinate of the point corresponding to an estimation error of the order of 10% (the second curve from below) is *r=N/T*= 0.025 which means that the length of the time series necessary to reach this level of error is 40 times longer than the number of assets in the portfolio.

The meaning of Δ is the following: Because of the assumed i.i.d. nature of the assets, the „true" optimal portfolio weights are all equal to 1/*N,* so their distribution is concentrated on this single value. However, we are observing the fluctuations of assets only for a period *T*, so the weights will deviate from *1/N* and follow a (Gaussian) distribution; Δ is proportional to the standard deviation of this distribution. Thus Δ measures the uncertainty of the weights within a given finite sample, but it also characterizes the fluctuations of weights from sample to sample.

The uppermost curve in Fig. 1 corresponds to $\Delta = \infty$. The distribution of weights is completely smeared out in the portfolios corresponding to the points along this curve, and the estimation error is infinitely large. This curve is the phase boundary: optimization of ES is not feasible above it. The curves below the phase boundary correspond to smaller and smaller estimation error. Note that all the finite $\Delta$ curves bend over and go to zero in the limit $\alpha \to 1$. Assume we have to work at a high confidence level, such as $\alpha = 0.975$, as proposed by the new market risk regulation. Assume furthermore that we wish to achieve an estimation error of 10%, that is $\Delta = 0.1$. The corresponding curve is the second from below. The $N/T$ ratio works out to be 0.025 at this point. This means that our time series should be 40 times longer than the dimension of the portfolio, in order to ensure a 10% estimation error at $\alpha = 0.975$. To illustrate the point, consider a portfolio of size, say, $N=100$: $T = 40N$ would amount to 16 years of daily returns. Larger portfolios and/or tighter error requirements would take even longer time series.

It has always been clear that downside risk measures depending on a small fraction of data demand a very large number of observations. Fig.1 and the theory behind it lends a quantitative meaning to this notion.

A word about the theory: One starts with noticing that the task of averaging over statistical samples is analogous to what is called „quenched avaraging" in the theory of random systems. One can therefore borrow the tools of this theory, in particular the method of replicas. Details about the derivation of the closed set of equations behind the results reported in Fig. 1 can be found in [4,5,8,10].

Regularization would, of course, completely modify the picture, and in [7,8,10] we also looked into the effect of various types of regularizers and their interpretation in terms of market impact [8,10]. Yet we refrain from including regularization in the present note, because we wish to show up how unstable the original, unregularized problem is, and how large the sample fluctuations in the resulting ES estimates are for any realistic values of the control parameters $N/T$ and $\alpha$. Such a strong instability in the original problem demands a strong regularizer to cure it, so strong indeed that an efficient regularizer will suppress not only the extreme fluctuations, but basically also all the information coming from the observations. Applying such a strong regularizer (as it were, a dominant Bayesian prior) would only mask rather than cure the disease: the optimal portfolio so obtained would reflect the structure of the regularizer rather than that of the data. In such a situation it may be better to forget about the observations altogether and rely on expert opinion – which is the practice in the case of most investment decisions anyhow.

Another apparent remedy is to use parametric estimates, instead of historical ones. We have checked this point and found that in the $\alpha \to 1$ limit all the parametric contour lines of ES (and also those of VaR) tend to 1 from below. This might raise the hope that the parametric estimation would be able to overcome the problem of insufficient data: the $N/T$ ratio required at the same confidence level (0.975) and the same 10% estimation error as used for the historical estimate is not small, it is above 0.5. This may suggest that the number of observations should be of the order of only twice the dimension of the portfolio. However, this is an illusion and, if many practitioners share it, a very dangerous one. In the parametric method one has to be able to reliably estimate the probability distribution of losses, especially in the asymptotic region where ES is to be measured. In real markets this probability distribution is *not* a Gaussian, but fat tailed, and estimating a power law like tail in the region of rare observations is as difficult as to obtain a reliable historical estimate.

There are two further points we wish to mention here. The analytic results on which Fig. 1 rests were derived for independent, identically distributed normal variables. Market fluctuations are neither independent, nor normal. In [5] we showed how correlations between normal variables can be accommodated within the replica formalism behind the results displayed in Fig. 1. At the expense of some slight additional complications we were able to consider any covariance matrix, and found that as long as the covariance structure is not too extreme and the covariance matrix is invertible, the moral of the tale remains the same.

As for the non-Gaussian nature of fluctuations, we do not have an analytic theory of the estimation error for an arbitrary distribution. We did, however, perform extended simulations and measured the estimation error numerically. The most important class to consider is that of the fat tailed distributions. Intuitively, it is clear that the fatter the tail the larger the estimation error will be. This is indeed born out by our simulations. We considered two kinds of fat tailed distributions: the Student t-distribution of degree of freedom $v=3$ (asymptotic behavior $1/x^4$) and the Cauchy distribution (with asymptotics $1/x^2$). (The former is more or less the asymptotic behavior of assets in the market, the latter is merely a mathematical example, no asset is known to fluctuate this strongly.) What we have been able to obtain so far are just the three phase boundaries: Somewhat surprisingly, the curves corresponding to the Gaussian, Student resp. Cauchy distributions are rather close to each other, in fact, it takes quite an effort to resolve them numerically. They all start at zero for α = 0, and all go to ½ for α = 1. For α's in between, they are slightly different, with the Gaussian phase bounday lying above the Student that, in turn, lies above the Cauchy curve. This means that in order to be able to solve the optimization problem at the same confidence level and the same portfolio size *N*, one needs the longest time series for Cauchy-distributed fluctuations, somewhat shorter for Student variables, and even shorter for the Gaussian variables. However, for α's of practical interest the three curves basically coincide. We believe that the finite Δ curves corresponding to theses three cases will display a similar ranking, but so far we have not been able to collect enough numerical data to substantiate this belief.

The message of this study is very simple: without a sufficient amount of data one cannot take an informed decision. What is new is the contour map of estimation error that allows one to calculate the expected value of the estimation error for Gaussian input variables, at a given confidence level and aspect ratio *N/T*. The lines of constant estimation error bending over and falling down to zero for α → 1 is a warning: the optimization of large portfolios at high confidence limits either requires an unrealistic amount of data, or leads to huge estimation errors.